\documentclass[twocolumn,aps]{revtex4}
\usepackage{graphicx}
\begin{document}
\title{Self-gravito-acoustic shock signals  in astrophysical compact objects}
 \author{A. A. Mamun} \affiliation{Department of Physics,
Jahangirnagar University, Savar, Dhaka-1342, Bangladesh}
\begin{abstract}
The existence  of self-gravito-acoustic (SGA) shock signals (SSs) associated with negative self-gravitational potential
in the perturbed state of  the astrophysical compact objects (ACOs)  (viz. white dwarfs, neutron stars, black holes, etc.) is  predicted for the first time.  A modified Burgers equation (MB), which is valid for both planar and  non-planar spherical geometries, by the reductive perturbation method.  It is shown that the longitudinal viscous force acting in the medium of any ACO  is the source of dissipation, and is responsible for the formation of these SGA SSs.  The  time evolution of these SGA SSs is also shown for different values (viz. $0.5$,  $1$, and $2$ ) of the ratio of nonlinear coefficient to dissipative coefficient in the MB equation. The theory presented here is so general that it can be applied in any ACO of planar or non-planar spherical shape.  
\end{abstract}
\maketitle
The astrophysical compact objects (ACOs)  (viz. white dwarfs, neutron stars, black holes, etc.) are  significantly  different from other terrestrial bodies not only because of their  extra-ordinarily high density and extremely low temperature \cite{Chandrasekhar1939,Shapiro2004,Koester1990,Koester2002,Horn1991}, but also because they can introduce new self-gravito-acoustic mode and associated new nonlinear structures.  They,  in fact, contain an
admixture of  degenerate, non-inertial particle species (viz.  electron or positron or non-zero mass \cite{Fraga2005} quark species, or any one/two or all of them),
non-degenerate or degenerate inertial light particle  species  (viz. proton or neutron or ${\rm ~^{4}_{2}He}$, 
 or ${\rm ~^{12}_{~6}C}$ or  ${\rm ~^{16}_{~8}O}$ species \cite{Shapiro2004,Koester1990,Koester2002},  or any one/two or all of them), and heavy particle species 
(viz. ${\rm ~^{56}_{ 26}Fe}$ or ${\rm ~^{85}_{ 37}Rb}$ or ${\rm ~^{96}_{42}Mo}$ species \cite{Vanderburg2015,Witze2014}, or any one/two or all of them).  

The degeneracy of non-inertial electron or positron or quark species arises due
to Heisenberg's uncertainty principle ($\Delta p\Delta x\ge
\hbar/2$, where $\hbar$ is the reduced Planck constant, $\Delta p$ is the uncertainty in the particle's momentum, and $\Delta x$ is the uncertainty in particle's position). This indicates
that the momentum of a highly compressed particle species is extremely
uncertain, since the particle species is located in an extremely confined
space. Therefore, even though this confined space is extremely cold, the 
particle species must move very fast on average, and give rise to a
very high pressure, known as `degenerate pressure', which depends only on degenerate particle number density. This means that in order to compress an object  into an extremely small space, a tremendous pressure,  which is the self-gravitational pressure in any ACO (viz. white dwarfs, neutron stars, black holes, etc. \cite{Shapiro2004,Chandrasekhar1939}) is required to
balance this degenerate pressure. 

Recent discovery \cite{Abbott2016} of gravitational waves
\cite{Abbott2016,Kurkela2016,Ho2016}  (produced by merging of two
black holes)  has motivated space and astrophysicists to search for new gravito-acoustic modes  that may exist  in such ACOs  (viz. white dwarfs, neutron stars, black holes, etc.).   The concept of a new self-gravito-acoustic (SGA) mode can be developed in a way that an equilibrium ACO  is disturbed by any of many reasons  (viz. merging \cite{Abbott2016} of two small ACOs, fragmentation \cite{Shapiro2004} of a large ACO, gravitational interaction
\cite{Shapiro2004} among neighboring ACOs,  etc.), and that if  the disturbed ACO is compressed (expanded),  the degenerate (self-gravitational) pressure  brings it back to its equilibrium shape, but during this action it is expanded (compressed) more than its equilibrium shape according to Newton's 1st law of motion, and  again the self-gravitational (degnerate pressure brings the system back to  its equilibrium shape, but again during this action, it is compressed (expanded) more than its equilibrium shape according to the same reason. These compression (rarefaction) and rarefaction (compression) of the system continue, and thus,
a new SGA mode is developed.  

The present article is aimed at identifying the  SGA shock signals associated with the self-gravitational (SG) potential in ACOs (viz.  white dwarfs, neutron stars, black holes, etc.) which are assumed to contain arbitrary number of non-inertial degenerate particle species $s$  (viz. electron
or/and positron or/and non-zero mass quark, etc. \cite{Shapiro2004,Horn1991}), and of inertial degenerate particle species $j$  (viz. proton or/and neutron, and $~^{4}_{2}$He or $~^{12}_{~6}$C or $~^{12}_{~6}$O, or/and $~^{56}_{26}$Fe or/and $~^{85}_{37}$Rd or/and $~^{96}_{42}$Mo, etc. \cite{Shapiro2004,Horn1991,Koester1990,Koester2002}). The perturbed state of such such ACOs can be described by generalized hydrodynamic model \cite{Ichimaru1986,Mamun2004,Haas2011,Shukla2011b,Mamun2011} in planar 
($\nu=0$) or nonplanar spherical ($\nu=2$) geometry  \cite{Maxon1974,Mamun2011} by 
\begin{eqnarray}
&&\hspace*{-8mm} \frac{\partial \psi}{\partial
r}=-\frac{3}{2}\alpha_s\frac{\partial \rho_s^{\frac{2}{3}}}{\partial r},
\label{be1}\\
&&\hspace*{-8mm}\partial_t\rho_j+\frac{1}{r^\nu}\partial_r(r^\nu\rho_ju_j) = 0,
\label{be2}\\
&&\hspace*{-8mm}(1+\tau_m d_t^j)\left[\rho_j\left(d_t^ju_j+\partial_r \psi+\frac{2}{3}\beta_j\partial_r\rho_j^{\frac{2}{3}}\right)\right]\nonumber\\
&&\hspace*{-1mm}=\frac{\eta}{r^\nu}\partial_r(r^\nu\partial_r u_j)
+\left(\zeta+\frac{\eta}{3}\right)\partial_r\left [\frac{1}{r^\nu}\partial_r (r^\nu u_j)\right],
\label{be3}\\
&&\hspace*{-8mm}\frac{1}{r^\nu}\partial_r(r^\nu\partial_r\psi)=\sum_s\delta_s\rho_s+\sum_j\mu_j\rho_j,
 \label{be4}
\end{eqnarray}
where $\partial_t= \partial/\partial t$,  $\partial_ r=\partial/\partial r$, and $d_t^j=\partial_t+u_j\partial_r$;  $\rho_s$ ($\rho_j$) is the number density of the degenerate,  non-inertial (inertial) particle species $s$ ($j$), and is normalized by
its equilibrium value $\rho_{s0}$ ($\rho_{j0}$);  
$u_j$ is the degenerate fluid speed of the species
$j$, and is normalized by $C_q$ in which $C_q=(\sqrt{\pi}\hbar
\rho_{e0}^{1/3}/m_e^{1/3}m_p)$,  $m_p$ ($m_e$) is the proton (electron) mass and $\rho_{e0}$  is the equilibrium  mass density of the electron species; $\psi$ is the self-gravitational potential, and is normalized by $C_q2$;  
$\alpha_s=(m_p/m_s)^2 (m_e/m_s)^{2/3}(\rho_{s0}/\rho_{e0})^{2/3}$ and
$\beta_j=(m_p/m_j)^2(m_e/m_j)^{2/3}(\rho_{j0}/\rho_{e0})^{2/3}$, 
in which $m_s$ ($m_j$) is the mass of the non-inertial (inertial) degenerate particle species $s$ ($j$); 
$t$ is the time variable normalized by
$\omega_{Jp}^{-1}=(4\pi G\rho_{p0})^{-1/2}$; $r$ is the space variable
normalized by $L_q=C_q/\omega_{Jp}$;  $\delta_s=\rho_{s0}/\rho_{p0}$
and $\mu_j=\rho_{j0}/\rho_{p0}$ in which $\rho_{p0}$ is the mass
density of the proton species;  $\eta$ ($\zeta$)  is the shear (bulk)  viscosity
coefficient, and is normalized by $\rho_{j0}C_qL_q$, and  $\tau_{m}$ is the viscoelastic relaxation time normalized by 
$\omega_{Jp}^{-1}$.  There are various approaches for calculating these transport coefficients, 
$\eta$, $ \zeta$, and $\tau_m$.  These have been widely discussed in the existing literature
 \cite{Ichimaru1986,Durisen1973,Shukla2011b,Mamun2011}.  We note that in (\ref{be1}) the non-inertial degenerate species $s$ are assumed to be non-relativistically degenerate.  This has been considered by many authors during the last few years  \cite{Manfredi2005,Shukla2006,Markcloud2007,Shukla2011a,MAS2016,Brodin2017,MAS2017}
to study the electro-acoustic or magneto-acoustic linear/nonlinear waves, but not to study any kind of self-gravito-acoustic waves/modes, which  is the basis of the present work.   We also note that (\ref{be1}) is obtained by equating  the outward degenerate pressure to the inward self-gravitational pressure of the  species $s$ \cite{Chandrasekhar1939}. This is, however, valid for the SGA perturbation mode whose phase speed is much
smaller than $C_{e}$, where $C_{e}=\sqrt{\pi}\hbar \rho_{e0}^{1/3}/m_e^{4/3}$. 

To construct a weakly nonlinear
theory  for the  nonlinear propagation of this perturbation mode by using the reductive perturbation method,  we,  first introduce the stretched co-ordinates \cite{Maxon1974,Shukla2011b,Mamun2011}:  $R=-\epsilon(r+V_0t)$,  $T=\epsilon^2t$ [where $V_0$  is normalized by $C_q$, and $R$ ($T$) is normalized by
$L_q$ ($\omega_{Jp}^{-1}$], and expand the perturbed quantities  in power series of $\epsilon$:
$\rho_{s,j}= 1+\epsilon\rho_{s,j}^{(1)}+\epsilon^2\rho_{j,s}^{(2)}+ \cdots$, 
$u_j=\epsilon u_j^{(1)}+\epsilon^2 u_j^{(2)}+\cdots$, and
$\psi=\epsilon\psi^{(1)}+\epsilon^2\psi^{(2)}+\cdots$, where  $\epsilon$ is an expansion
parameter ($0<\epsilon<1$).  We next  develop equations  in various powers of
$\epsilon$ by using the stretched co-ordinates, and the expansions of these perturbed quantities.  Now,  keeping  the
terms containing  $\epsilon^2$ from (\ref{be1})$-$ (\ref{be3}), and
 $\epsilon$ from (\ref{be4}),  we obtain a set of linear equations.  On the other hand,   keeping  the
terms containing  $\epsilon^3$ from (\ref{be1})$-$ (\ref{be3}), and
 $\epsilon^2$ from (\ref{be4}), we obtain a set of nonlinear equations. 
These linear and nonlinear sets of equations can be reduced to a modified Burgers (MB) equation in the form
 \begin{eqnarray}
\partial_T\psi^{(1)}+\frac{\nu\psi^{(1)}}{2T}-A \psi^{(1)}\partial_R
\psi^{(1)}= C \partial_R^2 \psi^{(1)},
\label{mBE}
\end{eqnarray}
where $A$ and $C$ are  the nonlinear and dissipation coefficients, respectively, and are given by
\begin{eqnarray}
&&\hspace*{-8mm}A=\frac{3}{2V_0\sum_j\mu_j\gamma_j^2}\left[\sum_j\mu_j\gamma_j^3\gamma_j^\prime+\frac{V_0^4}{9}\sum_s\delta_s\alpha_s^2\right], \\
&&\hspace*{-8mm}C=\frac{1}{2\sum_j\mu_j\gamma_j^2}\sum_j\mu_j\gamma_j^2\eta_j^l,
\end{eqnarray}
in which $\gamma_j=(1-\beta_j/V_0^2)^{-1}$, $\gamma_j^\prime=1+\beta_j/6V_0^2$,  and $\eta_j^l$  
$(=\zeta + 4\eta/3)$ is the longitudinal viscosity coefficient.   

\begin{figure*}[t!]
\centering
\includegraphics[width=15cm]{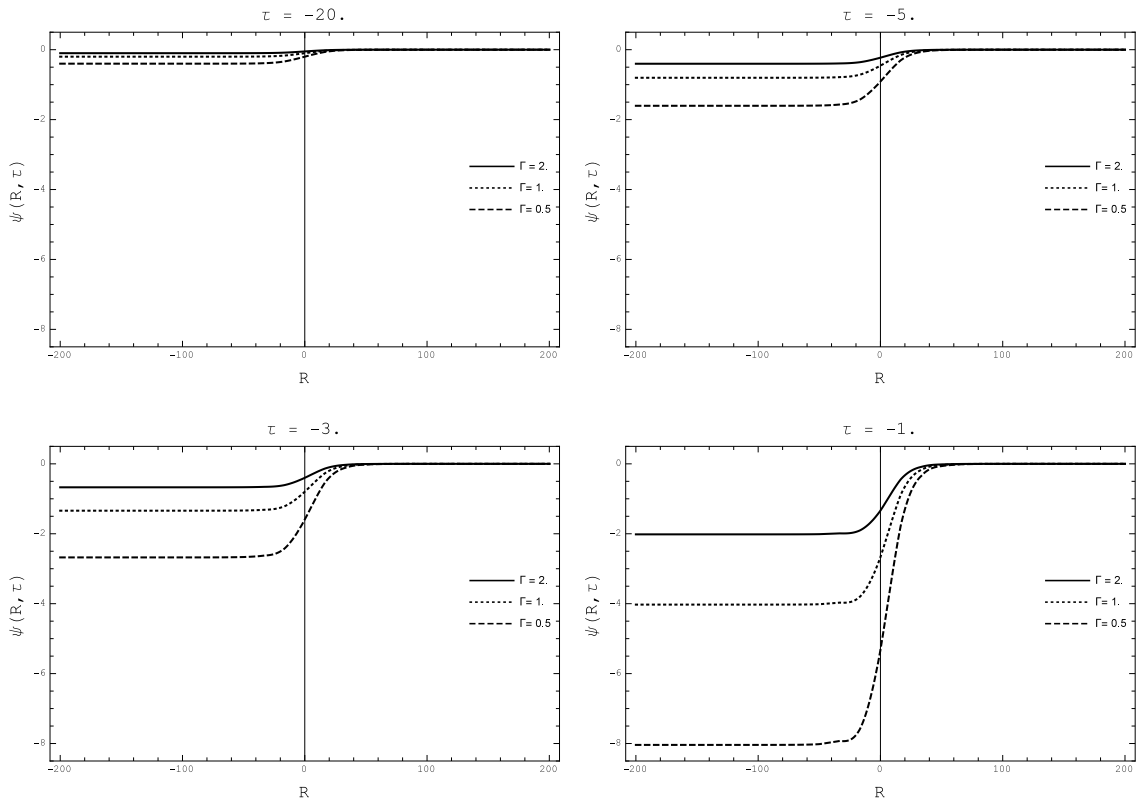}
\caption{The numerical solutions of (\ref{RmBE}) for different values of $\tau$ (viz. -20, -5, -1), and  $\Gamma$ (viz. 0.5, 1, and 2).  We choose  the initial pulse [represented by  (\ref{sol})] at an initial time $\tau=-20$, since at this intial time  the SGA SSs for the nonplanar (spherical)  geometry ($\nu=2$) are identical to those for the planar geometry ($\nu=0$).} 
\label{f1}
\end{figure*}

Now,  transforming $T$ to  $\tau=CT$,  and denoting $A/C$ by  $\Gamma$ (a ratio of the nonlinear coefficient $A$ to the dissipative coefficient $C$), one can  express the MB equation (\ref{mBE})  as
 \begin{eqnarray}
\partial_\tau\psi^{(1)}+\frac{\nu\psi^{(1)}}{2\tau}-\Gamma\psi^{(1)}\partial_R
\psi^{(1)}= \partial_R^2 \psi^{(1)}.
\label{RmBE}
\end{eqnarray}
It is obvious from this equation that the extra-term, $\nu\psi^{(1)}/2\tau$ is due to the effect of the non-planar spherical geometry [since this extra-term disappears for a planar geometry  ($\nu=0$)], and that the effect of this extra-term diminishes as $\tau$ become significantly large. This means that for a large value of $\tau$, the SGA SSs for the  nonplanar spherical spherical geometry ($\nu=2$) are identical to those for the planar geometry ($\nu=0$).  Thus, for a large value of $\tau$ or $\nu=0$, the stationary shock signal solution of (\ref{RmBE}) becomes \cite{Shukla2011b,Mamun2011}
\begin{eqnarray}
\psi^{(1)}=-\frac{1}{2}\psi^{(1)}_m\left[1-\tanh\left(\frac{\xi}{\Delta}\right)\right].
\label{sol}
\end{eqnarray}
where $\xi=R-U_0T$ (with $U_0$  being the speed of the frame of reference), and $\psi^{(1)}_m=2U_0/\Gamma$  ($\Delta=2/U_0$)  is  the height  (thickness) of the monotonic SGA SSs. This equation implies that the monotonic SGA SSs  are formed with $\psi<0$, since $\Gamma$ is numerical found to be positive for all possible values of the parameters 
 corresponding to the ACOs like white dwarfs and neutron stars.  To show how the SGA SSs evolve with time, and how they are significantly modified by $\Gamma$, the transformed  MB equation (\ref{RmBE})  is now numerically solved using an initial pulse represented by   (\ref{sol}).  The numerical results are displayed in figure \ref{f1}. 
which indicates that (i) for a large value of $\tau$ (e. g. -20)  planar and nonplanar  spherical SSs are identical for a fixed value of $\Gamma$ since for a large value of $\tau$  the extra term($ \psi^{(1)}/\tau$ for spherical geometry)  in (\ref{RmBE}) becomes insignificant ; ( ii) as the value of $\tau$ decreases, their height  and thickness of the SGA SSs increase. This is due to the effect of spherical geometry since for lower values of $\tau$ the extra term ($ \psi^{(1)}/\tau$) for spherical geometry)  in (\ref{RmBE}) becomes significant; and (iii) their height and thickness increase with  the decrease of  $\Gamma$.      

To summarize, a generalized hydrodynamic model has been used to treat  the nonlinear dynamics of different species of the SGDP systems like ACOs, which, in general, contains arbitrary number of degenerate, non-inertial particle  (viz. electron or/and positron or/and quarks) species,  and
arbitrary number of degenerate, inertial particle  (viz. proton or/and neutron or/and 
${\rm ~^{4}_{2}He}$ or/and ${\rm ~^{12}_{~6}C}$ or/and ${\rm ~^{56}_{ 26}Fe}$ or/and 
${\rm ~^{85}_{ 37}Rb}$ or/and ${\rm ~^{96}_{42}Mo}$).  The existence of  the SGA SSs  with $\psi<0$ in such a SGDP  system is  predicted for first time.  It is found here that for a large value of $\tau$  planar and nonplanar (spherical) SSs are identical, but they evolve with time significantly,  i. e.  the height  as well as the thickness of the SGA SSs increase as we observe them from an earlier time (viz. $\tau -20$)  to present time (viz. $\tau=-1$, since we cannot  observe them at $\tau=0$, where (\ref{RmBE}) has a pole).  It is also observed that the coefficient of longitudinal viscosity  ($\eta_j^l$) acts as a source of dissipation, and is responsible for the formation of the SGA SSs in the dissipative SGDP systems like ACOs, and that their height as well thickness  increases  with the increase in  the dissipative coefficient $C$, which is directly proportional to  $\eta_j^l$.     

The SGA SSs are associated with a new SGA mode in which  if a disturbed ACO is compressed (expanded),  the degenerate pressure  brings it back to its equilibrium shape, but during this action it is expanded (compressed) more than its equilibrium shape according to Newton's 1st law of motion, and  again the self-gravitational pressure brings the system back to  its equilibrium shape, but again during this action, it is compressed (expanded) more than its equilibrium shape according to the same reason, and so on.

The dissipative SGDP system considered here is generalized to arbitrary number of
non-inertial and inertial degenerate particle species with their
arbitrary mass densities. This theory is  also general from the point of view that it can be applied in any ACO, where the effect of the nonlinearity is comparable to or much less/more than that of the dissipation.  The investigation presented here can therefore be applied in any ACO. 

It should be noted here that (\ref{be4}) cannot be applied to
describe the equilibrium state of the SGDP system under consideration.
But this does not affect our investigation on the SGA SSs in
any  SGDQP under consideration. However, to explain the equilibrium
state of the SGDQP system under consideration, one has to rewrite the
basic equations in such a way that $n_{so}$, $n_{j0}$, and
$\psi_0$ (self-gravitational potential at equilibrium) are not equal to 
zero or constant, but function of $x$.  This means that $n_{s0}$, 
$n_{j0}$ and $\psi_0$ are
not constant, but are the function of $x$ so that the force
associated with degenerate pressures is balanced by the
self-gravitational force at equilibrium. This is the common
scenario for many ACOs like white dwarfs and neutron stars
\cite{Chandrasekhar1939}. To know the exact
variation of $\psi_0 (x)$ with $x$, one has to numerically solve
Poison's equation for $\psi_0 (x)$ by choosing the appropriate
variation of $n_{so}(x)$ and $n_{j0}(x)$ with $x$.  However, the results presented here (particularly,  concept of new SGA mode,   exitence of new SGA SSs  with new basic features (ploarity, hight, thikness, etc.)  are correct from both analytical and numerical points of view.\\
The author would like to thank Dr. A. Mannan for helping him during the numerical analysis (particularly, making the movie) of this work.

\end{document}